\documentclass[pdflatex,sn-mathphys-num]{sn-jnl}

\usepackage{graphicx}%
\usepackage{multirow}%
\usepackage{amsmath,amssymb,amsfonts}%
\usepackage{amsthm}%
\usepackage{mathrsfs}%
\usepackage[title]{appendix}%
\usepackage{xcolor}%
\usepackage{textcomp}%
\usepackage{manyfoot}%
\usepackage{booktabs}%
\usepackage{algorithm}%
\usepackage{algorithmicx}%
\usepackage{algpseudocode}%
\usepackage{listings}%
\usepackage{float}%
\usepackage{lineno}

\theoremstyle{thmstyleone}%
\theoremstyle{thmstyletwo}%

\theoremstyle{thmstylethree}%

\begin{document}


\title[Ion-temperature-gradient turbulence from finite to weak magnetic shear regime]{Ion-temperature-gradient turbulence from finite to weak magnetic shear regime}


\author[1]{\fnm{Zihao} \sur{Wang}}\email{wzh95@ustc.edu.cn}

\author[2]{\fnm{Tiannan} \sur{Wu}}\email{wutn@ustc.edu.cn}

\author*[1]{\fnm{Shaojie} \sur{Wang}}\email{wangsj@ustc.edu.cn}

\affil*[1]{\orgdiv{Department of Engineering and Applied Physics}, \orgname{University of Science and Technology of China}, \orgaddress{\street{Jinzhai Road No. 96}, \city{Hefei}, \postcode{230026}, \country{China}}}


\abstract{
The tokamak, a toroidal magnetic device confining a hot plasma, is one of the most advanced approaches to fusion energy.
A central obstacle is turbulent transport driven by the ion-temperature-gradient (ITG) mode.
Operation scenarios that rely on a weak or zero magnetic shear core, such as the ITER hybrid scenario~\cite{campbell2025introduction}, have achieved markedly improved confinement~\cite{kin2021dynamics}, yet the underlying mechanism has remained unclear.
Here we show that the radial width of ITG poloidal harmonics is constrained not only by the familiar parallel Landau damping, but also by the isotropy of micro-turbulence in the plane perpendicular to the magnetic field.
In the weak-shear limit, micro-isotropy dominates and constrains the radial width at one poloidal wavelength, yielding a critical magnetic shear $s_{\text{crit}} \approx 1/(2\pi)$.
Above this threshold, the extended Type II ballooning modes prevail; below it, the localized Type I modes emerge, comprising only two or three harmonics.
Global gyrokinetic simulations spanning DIII-D, JET~and ITER parameters confirm this geometric criterion.
Nonlinear simulations with sustained heating show that weak-shear plasmas spontaneously form internal transport barriers with the turbulence suppressed in the radial region determined by $|s| < s_{\text{crit}}$.
This geometric criterion defines the weak-magnetic-shear regime and offers a fresh perspective on turbulent transport in fusion plasmas.}

\maketitle

\section{Introduction}\label{sec1}

Fusion energy, if harnessed economically, promises to reshape the global energy landscape~\cite{ongena2012energy}.
The tokamak, a toroidal magnetic device confining a hot plasma (ionized gas), is one of the most advanced approaches to this goal~\cite{wesson2011tokamaks}.
To reach a commercially attractive reactor, however, the plasma must achieve high confinement and steady-state operation.
In many tokamaks, such as TFTR~\cite{levinton1995improved}, JT-60U~\cite{takenaga2009characteristics,kin2021dynamics}, JET~\cite{romanelli2010fast,mazzi2022enhanced}, ASDEX Upgrade~\cite{maggi2010pedestal}, DIII-D~\cite{greenfield1997transport}, EAST~\cite{gao2020experimental}, HL-2A~\cite{yu2016ion} and KSTAR~\cite{chung2021sustainable,han2022sustained}, high core plasma confinement has been demonstrated by forming an internal transport barrier (ITB)~\cite{connor2004review}, a localized region of steep pressure gradient due to the suppression of turbulence, such as the ion-temperature-gradient (ITG) mode~\cite{liewer1985measurements,dimits1996scalings,horton1999drift}.
Externally driven $E\times B$ flow shear~\cite{biglari1990influence,burrell1997effects}, zonal flows~\cite{lin1998turbulent}, reversed magnetic shear~\cite{levinton1995improved}, the Shafranov shift~\cite{synakowski1997roles}, energetic-particle effects~\cite{han2022sustained,mazzi2022enhanced,citrin2023overview} and weak magnetic shear~\cite{greenfield1997transport,kin2021dynamics} are important in turbulence suppression.
The magnetic shear is $s = (r/q)\mathrm{d}q/\mathrm{d}r$, with $q$ the safety factor and $r$ the minor radius.
The ITB formed in the weak-magnetic-shear configuration is of particular importance, since this configuration has been designed in the International Thermonuclear Experimental Reactor (ITER) with the hybrid operation scenario~\cite{shimada2007chapter,campbell2025introduction}.

The ITG mode in a tokamak exhibits a ballooning structure: the perturbation is unstable on the bad-curvature region and stable on the good-curvature region~\cite{hirose1995ion,kishimoto1999toroidal}.
Under finite magnetic shear, rational surfaces, those on which the safety factor $q$ equals a rational number, are closely spaced, and toroidal coupling binds many poloidal harmonics into a radially extended eigenmode known as the Type II ballooning mode~\cite{mercier1979plasma}, which can be treated with the ballooning transform~\cite{connor1979high,romanelli1989ion,guo1993linear}.
When the shear becomes weak, however, the rational surfaces become sparse, and the ballooning transform breaks down~\cite{connor2004microstability,candy2004smoothness}, and a critical shear was predicted by using a fluid model~\cite{connor2004microstability}.
The non-standard ballooning structure was observed in global gyrokinetic simulations~\cite{zielinski2020global,zhao2021linear}, but the predicted critical shear~\cite{connor2004microstability} is much larger than simulations suggest~\cite{wu2026effects}.
To date, no simple criterion exists for the threshold of the transition from extended to localized ITG structures -- a question left open by previous studies and a key gap in the predictive capability required for future reactors.

In a tokamak, charged particles stream freely along the magnetic field line but are constrained in the plane perpendicular to the field line.
An unstable mode~such as the ITG mode is anisotropic in three dimensions: its parallel wavelength is much larger than its perpendicular one.
In the perpendicular plane, since the poloidal wavelength $\lambda_\theta$ of the unstable mode is much smaller than the minor radius, the equilibrium is rotationally invariant on the micro-scale set by $\lambda_\theta$.
Due to this rotational invariance, we propose that the vortex structure of each poloidal harmonic must be isotropic in the perpendicular plane: its radial extent is comparable to its poloidal wavelength.
In the weak-shear regime, the radial width of a harmonic, $\Delta_{\text{mph}}$, is comparable to $\lambda_\theta$.
This micro-isotropy constraint, combined with the familiar parallel Landau damping constraint, yields a geometric critical magnetic shear $s_{\text{crit}} \approx 1/(2\pi)$,
which is insensitive to device size, safety factor, temperature, and density profiles.
When $|s| > s_{\text{crit}}$, the structrue of the ballooning mode is the extended Type II; when $|s| < s_{\text{crit}}$, the structure is the localized Type I consisting of only two or three poloidal harmonics.
We have verified this prediction by using the first-principle global gyrokinetic simulations across DIII-D, JET~and ITER parameters.

In nonlinear gyrokinetic simulations with sustained heating, we find that plasmas with weak magnetic shear exhibit substantially weaker ITG turbulence in the radial region with $|s| < s_{\text{crit}}$, achieve higher confinement, and spontaneously form stronger ITBs than their finite-shear counterparts.
These simulations provide a direct link between the geometric threshold $s_{\text{crit}}$ and the improved confinement observed in weak-magnetic-shear regimes.
Our findings suggest that the micro-isotropy constraint plays a fundamental and previously unrecognized role in magnetic confinement, offering a fresh perspective on turbulent transport in fusion plasmas and potentially guiding the design and operation of future tokamak fusion reactors.

\section{Results}\label{sec2}

An eigenmode with given toroidal mode number $n$ in a tokamak is composed of poloidal harmonics, each labelled by a poloidal mode number $m$ and centred at a radial location $r_m$.

\textbf{Parallel Landau damping (PLD) constraint.}
It is familiar that the weak PLD, $|k_{\parallel}|v_{\text{th,i}} < \alpha_{\text{PLD}}|\omega_r|$, is required for instabilities to develop~\cite{hasegawa1975plasma}.
$k_\parallel = (m-nq)/(qR)$ is the parallel wavenumber of the $m$-th harmonic, with $\omega_r$ the mode frequency, $v_{\text{th,i}}$ the ion thermal velocity and $R$ the major radius.
$\alpha_{\text{PLD}} \sim 1$.
Each harmonic must stay close to its rational surface $r_{s,m}$, defined by $q(r_{s,m}) = m/n$.
The PLD constraint imposes a limit on the radial width of a single harmonic $\Delta_{\text{mph}} < \Delta_{\text{PLD}}$.
\begin{equation} \label{eq:constraint_1}
    \Delta_{\text{PLD}} = 2|m-nq(r)|\cdot \Delta_{\text{mrs}} = \frac{2\alpha_{\text{PLD}}}{k_\theta |s|} \frac{|\omega_r|}{\omega_{\text{t}}},
\end{equation}
with $\Delta_{\text{mrs}} = r/(nq|s|) = 1/(k_\theta |s|)$ the spacing of adjacent rational surfaces, $\omega_{\text{t}} = v_{\text{th,i}}/(qR)$ the ion transit frequency and $k_\theta \approx nq/r$ the poloidal wavenumber.

\textbf{Micro-isotropy constraint.}
The micro-isotropy constraint on the harmonic radial width is
\begin{equation} \label{eq:constraint_2}
    \Delta_{\text{iso}} = \alpha_{\text{iso}}\lambda_{\theta},
\end{equation}
with $\alpha_{\text{iso}} \sim 1$.

Therefore, $\Delta_\text{mph}$ is subject to both the PLD and the micro-isotropy constraints.
The actual width is the smaller of the two:
\begin{equation} \label{eq:constraint}
    \Delta_{\text{mph}} = {\text{min}}\left\{ \Delta_{\text{PLD}}, \Delta_{\text{iso}} \right\}.
\end{equation}
$\Delta_{\text{PLD}} \propto 1/|s|$ grows without bound as $|s| \to 0$, while $\Delta_{\text{iso}}$ is shear-independent.
The two bounds cross at $|s|=\hat{\omega}$, with the normalized frequency $\hat{\omega} \equiv (\alpha_{\text{PLD}}|\omega_r|)/(\alpha_{\text{iso}} \pi \omega_{\text{t}})$.
For $|s| > \hat{\omega}$, the PLD constraint dominates; for $|s| < \hat{\omega}$, the isotropy constraint dominates.

\begin{figure}
    \centering
    \includegraphics[scale = 1]{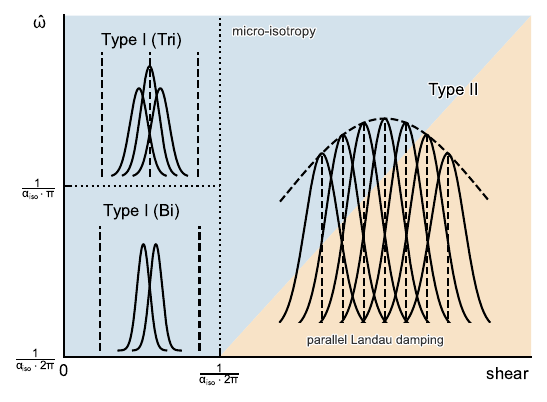}
    \caption{\textbf{Transition of ITG ballooning modes.} Two constraints compete to set the radial width of a poloidal harmonic, and the background marks which one dominates: micro-isotropy above the diagonal $|s|=\hat{\omega}$ and PLD below it, where $\hat{\omega}$ is the normalized frequency plotted on the vertical axis. The vertical dotted line marks the critical shear $|s|=1/(2\pi\alpha_{\text{iso}})$, the horizontal dotted line is the Bi--Tri boundary. The three panels depict the characteristic radial structures of Type II, Type I (Tri), and Type I (Bi) modes, placed in the corresponding parameter regions. The vertical dashed lines denote the rational surfaces.}
    \label{fig:ballooning_mode}
\end{figure}

\textbf{Critical magnetic shear.}
To form the ballooning structure, a poloidal harmonic must be toroidally coupled to its sideband harmonics, which requires that they overlap radially.
Since the spacing between adjacent harmonics is $|r_{m+1} - r_m|$~\cite{connor2004microstability}, the overlap condition reads $\Delta_\text{mph} > |r_{m+1}-r_m|$.

When the magnetic shear is finite, the constraint on $\Delta_{\text{mph}}$ and the radial overlap condition are met in a straightforward way.
The ballooning transform that describes Type II mode invokes the translational invariance along the radial direction~\cite{connor1978shear}.
Each harmonic sits on its own rational surface, so that $|r_{m+1}-r_m| = \Delta_{\text{mrs}}$.
For $|s| > \hat{\omega}$, $\Delta_{\text{mph}} = \Delta_{\text{PLD}}$; in this regime, since both $\Delta_{\text{PLD}}$ and $\Delta_{\text{mrs}}$ scale as $1/(k_\theta |s|)$, the overlap condition reduces to $\Delta_{\text{PLD}} > \Delta_{\text{mrs}}$, which is shear-independent.
Toroidal coupling binds many harmonics into the familiar radially extended Type II ballooning mode~\cite{mercier1979plasma}.

When $|s| < \hat{\omega}$, the micro-isotropy constraint dominates ($\Delta_{\text{mph}} = \Delta_{\text{iso}}$), while $\Delta_{\text{mrs}}$ continues to grow.
The overlap condition becomes $\Delta_{\text{iso}} = \Delta_{\text{mrs}}$,
where the factor $k_\theta$ cancels, yielding the critical magnetic shear
\begin{equation}
    s_{\text{crit}} \equiv \frac{1}{2\pi\alpha_{\text{iso}}} \approx \frac{1}{2\pi}.
\end{equation}
Here we used $\alpha_{\text{iso}} \approx 1$, which is justified later.
The Type II arrangement $\Delta_{\text{mph}} > \Delta_{\text{mrs}}$ is satisfied here only if $|s| > s_{\text{crit}}$.

When $|s| < s_{\text{crit}}$, $\Delta_{\text{mph}} < \Delta_{\text{mrs}}$.
Therefore, Type II modes can no longer satisfy the overlap condition, and the eigenmode collapses into a localized structure, the Type I ballooning mode~\cite{mercier1979plasma}.
The Type I mode satisfies the overlap condition not by increasing $\Delta_{\text{mph}}$, but by shrinking the effective spacing: all participating harmonics share a common localization centre, so that $|r_{m+1} - r_m| \ll \Delta_{\text{mrs}}$.

Two subtypes of Type I exist, distinguished by how they satisfy the PLD constraint while sharing a common centre.
For Type I (Tri-harmonic), three harmonics $(m_0, m_0\pm 1)$ centred at $r_{s,m_0}$. The sidebands have $|m-nq(r_m)| \approx 1$ with $m=m_0\pm 1$. The PLD constraint gives $\alpha_{\text{PLD}}\omega_r/\omega_{\text{t}} > 1$.
For Type I (Bi-harmonic), two harmonics $(m_0, m_0+1)$ centred on the half-rational surface where $|m_0-nq(r_{m_0})| = |m_0+1-nq(r_{m_0+1})| = 1/2$. Both harmonics have $|m-nq(r_m)| \approx 1/2$. The PLD constraint gives $\alpha_{\text{PLD}}\omega_r/\omega_{\text{t}} > 1/2$.
As $\omega_r$ increases with the ITG mode drive, the system crosses from Bi to Tri at $\omega_r/\omega_{\text{t}} > 1/\alpha_{\text{PLD}}$;
this Bi--Tri transition is used to calibrate $\alpha_{\text{PLD}}$, which gives $\alpha_{\text{PLD}} \approx 2$ (Extended Data Fig.~\ref{fig:alpha_PLD}).

The transition from the finite- to the weak-shear regime is shown schematically in Fig.~\ref{fig:ballooning_mode}.


We verify these predictions by using the first-principle global gyrokinetic code NLT~\cite{ye2016gyrokinetic,xu2026numerical}.
Three devices (DIII-D~\cite{luxon2005brief}, JET~\cite{rebut1985joint} and ITER~\cite{aymar2002iter}) are examined with Cyclone-like and flat temperature profiles (Methods).

\begin{figure}
    \centering
    \includegraphics[scale = 1]{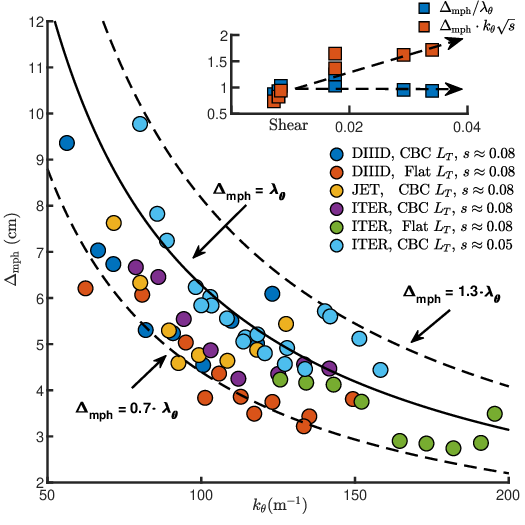}
    \caption{\textbf{Numerical calibration of $\alpha_{\text{iso}}$.} Measured full width at half maximum $\Delta_{\text{mph}}$ for Type I (Bi) modes in DIII-D, JET~and ITER. Each point is one simulation; two dotted lines mark $\alpha_\text{iso} = 1.0 \pm 0.3$. Inset, $\Delta_{\text{mph}}/\lambda_\theta$ versus the local magnetic shear for a dedicated set of weak-shear equilibria; the range $|s| \leq 0.04$ is well below $s_{\text{crit}}$. The horizontal line marks $\alpha_{\text{iso}} = 1$. The quantity $\Delta_{\text{mph}} k_\theta\sqrt{|s|}$, expected to be constant under the fluid $|s|^{-1/2}$ scaling~\cite{connor2004microstability}, but is shown to increase with $|s|$ by three times.}
    \label{fig:calibration}
\end{figure}

\textbf{Test of micro-isotropy.} We measure the radial width of Type I (Bi) modes in all three devices.
Figure~\ref{fig:calibration} shows $\Delta_{\text{mph}}/\lambda_\theta$ for a wide range of cases and suggests that $\alpha_{\text{iso}} = 1.0 \pm 0.3$.
A central prediction of the theory is that, when the micro-isotropy constraint dominates ($|s| < \hat{\omega}$), the harmonic width becomes independent of magnetic shear, $\Delta_{\text{mph}} \approx \lambda_\theta$.
This stands in clear contrast to the scaling $\Delta_{\text{mph}} \propto 1/(k_\theta\sqrt{|s|})$ obtained previously in Ref.~\cite{connor2004microstability}.
The inset of Fig.~\ref{fig:calibration} tests this directly, by using a dedicated set of weak-shear equilibria that span $|s| \leq 0.04$, well inside the weak-shear regime ($|s| \ll s_{\text{crit}}$).
$\Delta_{\text{mph}} \cdot k_\theta\sqrt{|s|}$ would rise as $|s|$ decreases, by a factor of three over the range probed here; however, $\Delta_{\text{mph}}/\lambda_\theta$ is almost constant with $\alpha_{\text{iso}} = 1.0 \pm 0.1$ as $|s| \to 0$.
Therefore, one concludes that the harmonic radial width is independent of shear in the weak shear regime.
The linear and nonlinear tests of $s_{\text{crit}}$ in the following further support $\alpha_{\text{iso}} \approx 1$.

\begin{figure}[h]
    \centering
    \includegraphics[scale=1]{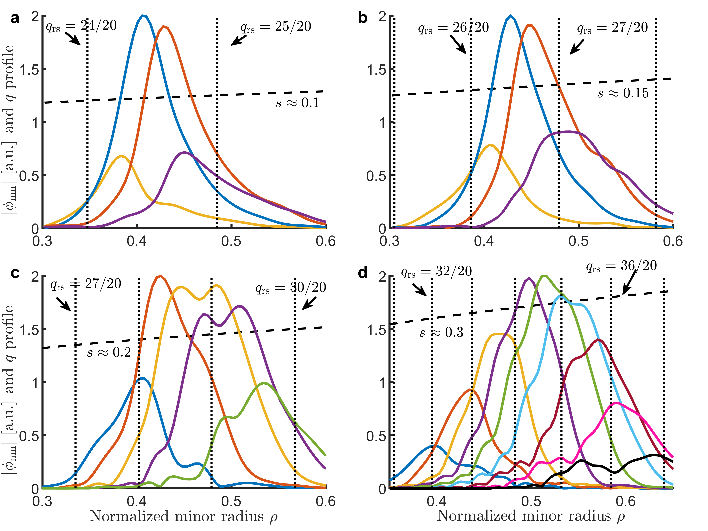}
    \caption{\textbf{Toroidal coupling structures across the critical shear.} \textbf{a-d}, Poloidal harmonic amplitudes and $q$ profiles as a function of normalized minor radius $\rho=r/a$ for four values of magnetic shear: (\textbf{a}) $s=0.1$, (\textbf{b}) $s=0.15$, (\textbf{c}) $s=0.2$, (\textbf{d}) $s=0.3$. All cases use DIII-D Cyclone-like profiles (Methods). Vertical black dashed lines indicate rational surfaces. At $|s|\le 0.15$, the eigenmode is Type I (Bi); for $|s|\ge 0.2$, it is Type II. The transition takes place at $s \approx 0.16$, which is consistent with $s_{\text{crit}} \approx 1/(2\pi)$, much less than $0.5$ predicted by Ref.~\cite{connor2004review}.}
    \label{fig:s_threshold}
\end{figure}

\textbf{Linear test of $s_{\text{crit}}$.}
Figure~\ref{fig:s_threshold} displays the toroidal coupling structure of ITG eigenmodes for a DIII-D Cyclone-like case at four shear values.
At $s = 0.1$ and $0.15$, the eigenmode is of Type I.
At $s = 0.2$ and $0.3$, the mode spans multiple rational surfaces, characteristic of Type II.
The structural transition thus lies in the interval $0.15 < s_{\text{crit}} < 0.20$.
The same behaviour is observed across other devices and equilibrium profiles.
For JET and ITER parameters, eigenmodes with $|s| \leq 0.15$ are invariably of Type I, while those with $|s| \geq 0.20$ are invariably of Type II; a weakly reversed core shear follows the same classification (Extended Data Fig.~\ref{fig:linear_reversed}).
This insensitivity to device size and equilibrium supports the geometric origin of the shear threshold.

To describe the transition of the ITG mode from the finite- to the weak-shear regime, we have introduced rotational invariance as the underlying principle, which leads to micro-isotropy of the harmonic vortex.
This micro-isotropy is a property of the linear eigenmode and is therefore distinct from the statistical micro-isotropy of nonlinear turbulence in neutral fluids~\cite{kolmogorov1941local}.
In the same spirit, translational invariance underlies the ballooning transform used to describe the Type II ballooning mode~\cite{connor2004microstability,dewar1983ballooning}.
As the magnetic shear increases, rotational invariance breaks down and the familiar translational invariance takes over, so that the ballooning mode changes from Type I in the weak-shear regime to Type II in the finite-shear regime.

\begin{figure}[H]
    \centering
    \includegraphics[width=\textwidth]{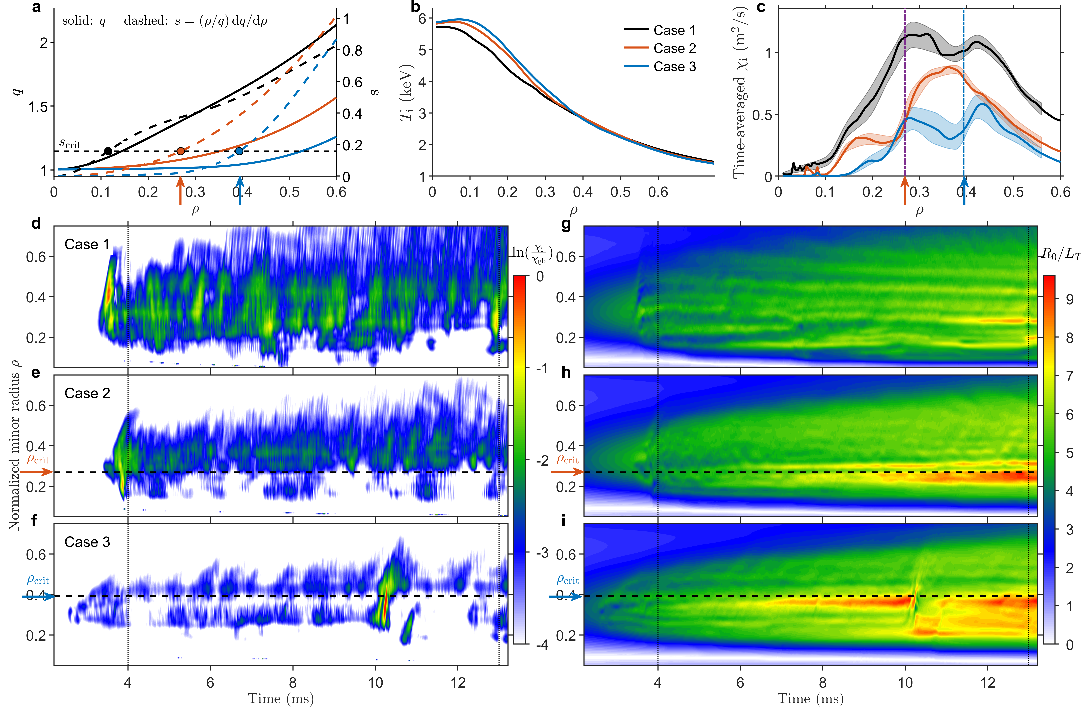}
    \caption{\textbf{Nonlinear ITG turbulence simulation in finite- and weak-shear configurations.} \textbf{a}, Safety factor $q$ (solid, left axis) and magnetic shear $s=(\rho/q)\,\mathrm{d}q/\mathrm{d}\rho$ (dashed, right axis) versus normalized minor radius $\rho=r/a$ for the three equilibria: Case~1 (black), Case~2 (orange) and Case~3 (blue). The horizontal dashed line is $s_{\text{crit}}=1/(2\pi)$. The filled circles and the arrows below the axis mark $\rho_{\text{crit}}$, the radius at which each shear profile crosses $s_{\text{crit}}$. The three equilibria share the same initial density, temperature and heating profiles and differ only in the core safety-factor profile. \textbf{b}, Ion temperature profiles at the end of the simulation; $T_{\text{i}}(\rho=0.2) \approx 4.4$, $4.8$ and $5.0~\text{keV}$ in Cases 1, 2 and 3, respectively. \textbf{c}, Time-averaged (4--13~ms) ion heat diffusivity $\chi_{\text{i}}$, whose core radial-averaged reaches approximately $1.0$, $0.6$ and $0.4~\text{m}^2/\text{s}$ in Cases 1, 2 and 3, respectively; shadow areas denote the standard deviation over the averaging window (5000 time samples). \textbf{d-f}, Spatiotemporal evolution of the logarithmic normalized ion heat diffusivity $\text{ln}(\chi_{\text{i}}/\chi_{\text{gb}})$ in $\rho$ and time for Cases~1--3, and \textbf{g-i}, the corresponding normalized ion temperature gradient $R_{0}/L_{T}$. The horizontal dashed lines drawn in \textbf{e-f, h-i} mark the corresponding $\rho_{\text{crit}}$. In Cases~2 and 3 the turbulence is suppressed inside $\rho_{\text{crit}}$ and a steep-gradient region (ITB) develops. The location of a ITB foot is identified as the place where $R_{0}/L_{T}$ begins to significantly increase; in Case~1 turbulence reaches the core and no ITB forms. Here $\chi_{\text{gb}}=\rho_{\text{i}}^{2}c_{\text{s}}/a \approx 11~\text{m}^2/\text{s}$ is the gyro-Bohm diffusivity, with $\rho_{\text{i}}$ the sound gyroradius, and $L_T=T_{\text{i}}/|\nabla T_{\text{i}}|$. The orange and blue arrows denote the location of $\rho_{\text{crit}}$.}
    \label{fig:nonlinear}
\end{figure}

\textbf{Nonlinear test of $s_{\text{crit}}$.}
We performed long-time ($13~\text{ms}$) nonlinear gyrokinetic simulations with sustained external heating using the NLT code with the Neighbouring Equilibrium Update method~\cite{wang2024self}, for three equilibria that share the same density, temperature and heating profiles and differ only in the core safety-factor profile (Extended Data Fig.~\ref{fig:profile_nonlinear}).
Zonal flows are generated self-consistently by the turbulence in these simulations, with no externally imposed $E\times B$ flow shear.
The three cases have different widths of the weak-magnetic-shear region, which is characterised by $\rho_{\text{crit}}$, the radius at which the shear profile crosses $s_{\text{crit}}$: $\rho_{\text{crit}} \approx 0.10$, $0.27$ and $0.40$ for Cases 1, 2 and 3, respectively (Fig.~\ref{fig:nonlinear}a).
Figure~\ref{fig:nonlinear}(d-i) shows the spatiotemporal evolution of the ion heat diffusivity $\chi_\text{i}$ and the normalized ion temperature gradient $R_0/L_T$.

An ITB is identified when $R_0/L_T$ is large (Fig.~\ref{fig:nonlinear}(h, i)) due to turbulence suppression (Fig.~\ref{fig:nonlinear}(e, f)) in a radially localized region with $|s| < s_{\text{crit}}$.
The finite-shear Case 1 exhibits strong turbulence that reaches the plasma core down to $\rho \approx 0.1$ and a moderate temperature gradient with no ITB.
In Cases 2 and 3, whose weak-shear regions are radially wider, the turbulence recedes from the core and a steep gradient region develops.
The foot of the barrier falls at $\rho_{\text{crit}}$ (Fig.~\ref{fig:nonlinear}(h, i)).
The time-averaged core $\chi_\text{i}$ decreases with $\rho_{\text{crit}}$ (Fig.~\ref{fig:nonlinear}(e, f)).
The intermittent burst near $t \simeq 11$~ms in Case 3 is an avalanche of the subcritical turbulence~\cite{wang2024self}; it is transient and does not appreciably change the time-averaged transport level.

The physical interpretation is straightforward.
In the weak-shear core, the micro-isotropy constraint prevents the formation of extended Type II modes.
Localized Type I modes survive, but being radially narrow, they generate a short radial correlation length and hence a lower transport level~\cite{connor2004microstability}, so that they saturate at a steeper temperature gradient.
This is borne out in the saturated state, where the finite-shear run shows a longer radial correlation length than the weak-shear runs (Extended Data Fig.~\ref{fig:nonlinear_structure}).
In all runs the toroidal spectrum becomes broadband once the turbulence saturates, so the distinction lies in the radial structure rather than in the toroidal mode number.
Zonal flows develop self-consistently in all runs; what differs is the harmonic structure set by the core shear (Extended Data Fig.~\ref{fig:nonlinear_structure}).

\section{Conclusion}
The radial width of an ITG poloidal harmonic, $\Delta_{\text{mph}}$, is constrained by two competing constraints: the familiar PLD, and the proposed micro-isotropy in the plane perpendicular to the magnetic field.
In the finite shear regime, $\Delta_{\text{mph}}$ is constrained by the PLD, while in the weak shear regime, $\Delta_{\text{mph}}$ is constrained by the micro-isotropy.
The critical shear is geometric, $s_{\text{crit}} \approx 1/(2\pi)$, independent of device size, safety-factor profile and toroidal mode number.
This criterion holds across DIII-D, JET and ITER equilibria, and for reversed as well as monotonic shear.
The nonlinear simulations show that weak-shear core exhibits significantly reduced turbulent transport, a steepened temperature gradient, and the spontaneous formation of an ITB with the barrier foot at the normalized radius $\rho_{\text{crit}}$ where the equilibrium shear profile crosses $s_{\text{crit}}$.

Experimental observations of ITB formation in weak-shear core plasmas on a number of devices~\cite{takenaga2009characteristics,kin2021dynamics,romanelli2010fast,maggi2010pedestal,greenfield1997transport,gao2020experimental,chung2021sustainable,deng2022investigation} are consistent with our conclusion;
for example, from Figs. 1 and 2 in Ref.~\cite{deng2022investigation}, one concludes that an ion ITB is formed in the core region where $|s|<s_{\text{crit}}$.
The criterion may not be specific to the electrostatic ITG mode.
The kinetic ballooning and Alfv\'{e}nic ITG modes in the weak shear regime have also been revealed as radially localized modes;
We note that the mode structures reported in Fig. 4 of Ref.~\cite{zhao2021linear} and in Fig. 4 of Ref.~\cite{li2026gyrokinetic} show that the harmonic radial width is much smaller than the spacing of adjacent rational surfaces and is close to the poloidal wavelength, so they also satisfy the micro-isotropy constraint proposed here.

The present simulations employ the electrostatic ITG model with adiabatic electrons in tokamaks with circular cross section; kinetic electron effects, electromagnetic fluctuations, plasma shaping may modify the quantitative results.
Nonetheless, the agreement between the shear criterion and gyrokinetic simulations across three devices of vastly different size is striking.
It suggests that $s_{\text{crit}}$ could serve as a practical criterion for scenario design, helping to ensure that future fusion reactors operate in the parameter regime with significantly reduced turbulence and improved confinement of core plasmas.

\section{Methods}\label{sec11}

\subsection{NLT code.}
The NLT code is a nonlinear global gyrokinetic continuum code that solves the gyrokinetic Vlasov equation using the Numerical Lie-Transform (NLT) method~\cite{wang2012transport,wang2013kinetic,wang2013nonlinear,ye2016gyrokinetic,xu2017nonlinear}.
The NLT code has been benchmarked against GENE, ORB5 and other codes for linear ITG modes and nonlinear ITG turbulence~\cite{zhao2021time,zhang2024linear}, and has been validated against experimental results on EAST and HL-2A~\cite{hu2023gyrokinetic,yang2024verification}.

The NLT code uses the $\delta f$ scheme, which is efficient at low fluctuation levels but breaks down when the perturbation becomes comparable to the equilibrium in long-time simulations.
To overcome this limitation, the Neighbouring Equilibrium Update (NEU) method has been developed~\cite{wang2024self}.
When the fluctuation becomes too large, NEU re-partitions the system: the zonal components of the electrostatic potential and the distribution function are moved from the perturbation into the equilibrium, with the updated equilibrium distribution function satisfying the constant-of-motion condition $\{F_0, H_0\} = 0$.
This operation leaves the total system unchanged but keeps the perturbed level sufficiently low.
This capability is exploited in the nonlinear simulations reported in the main text.

\subsection{Physical model and simulation settings.}
The simulations employ the electrostatic gyrokinetic model with adiabatic electrons and concentric circular flux-surface equilibria.
The equilibrium profiles, including the safety factor $q(r)$, are held fixed.
The polarization density is computed using the double gyro-average method, making the code applicable to ITG modes of arbitrary perpendicular wavelength.
The simulation domain includes the magnetic axis~\cite{wu2026finite}.
Field-aligned coordinates $(\psi, \alpha, \theta)$ are used, with $\alpha = q\theta - \zeta$.
For all simulations, $n_{\text{ga}} = 8$ gyro-averaging points are used.
The grid parameters are $N_\psi \times N_\alpha \times N_\theta \times N_{v_\parallel} \times N_\mu = N_\psi \times 190 \times 16 \times 96 \times 16$.
The simulation domain spans $\rho = r/a \in [0, 1]$, $\alpha \in [0, 2\pi/n_{\text{tm}}]$, $\theta \in [0, 2\pi]$, $v_\parallel/v_\text{th,i} \in [-4.2, 4.2]$, and $\mu B_0/T_{\text{ref}} \in [0, 25.5]$, where $v_{\text{th,i}} = \sqrt{2T_{\text{ref}}/m_\text{i}}$ is the ion thermal velocity, $c_{\text{s}} = \sqrt{T_{\text{i}}/m_\text{i}}$ the ion sound speed and $\rho_{\text{i}} = c_{\text{s}}/\Omega_{\text{i}}$ the sound gyroradius used to normalize lengths and diffusivities.
Device-specific parameters (radial grid number $N_\psi$, toroidal sector $n_{\text{tm}}$, major radius $R_0$, minor radius $a$, on-axis magnetic field $B_0$, reference density $N_{\text{ref}}$, reference temperature $T_{\text{ref}}$) are summarized in Table~\ref{tab:parameters_setting}.
Convergence tests were performed by doubling $N_\theta$ (16 to 32), and by increasing $N_{v_\parallel}$ (96 to 128), and by halving the time step $\Delta t\cdot(\omega_{\text{i}}/2\pi)$ (4 to 2).
The results for the volume-averaged ion heat flux, the zonal radial electric field and the ion temperature profile showed no discernible difference among these tests.
For the same $k_\theta$, the unstable toroidal mode number $n$ depends on the devices; therefore the agreement of $s_{\text{crit}}$ across DIII-D ($n= 10 - 30$), JET ($n= 32 - 88$) and ITER ($n= 50 - 150$), whose dominant $n$ differ substantially, itself demonstrates the insensitivity to the toroidal mode number.

\begin{table}[ht]
    \centering
    \begin{tabular}{c|c|c|c|c|c|c|c}
        \hline
        Device & $N_\psi$ & $n_{\rm tm}$ & $R_0(\rm m)$ & $a(\rm m)$ & $B_0(\rm T)$ & $N_{\rm ref}({\rm m}^{-3})$ & $T_{\rm ref}(\rm keV)$ \\
        \hline
        DIII-D & 220 & 2 & 1.67 & 0.60 & 1.90 & $1.0\times10^{19}$ & 1.97 \\
        JET    & 375 & 4 & 2.96 & 1.25 & 3.45 & $3.0\times10^{19}$ & 5.0  \\
        ITER   & 625 & 8 & 6.20 & 2.00 & 5.30 & $8.0\times10^{19}$ & 12.0 \\
        \hline
    \end{tabular}
    \caption{\textbf{Simulation parameter settings across devices}}
    \label{tab:parameters_setting}
\end{table}

\subsection{Equilibrium profiles in linear simulations.}
Two sets of equilibrium profiles are used in the linear simulations. The Cyclone~base case (CBC) profiles~\cite{dimits2000comparisons} are defined by
\begin{equation*}
\begin{aligned}
    T_0(r) &= T_{\text{ref}} \exp[-\kappa_T \frac{a}{R_0} \Delta_T \tanh(\frac{(r-r_0)/a}{\Delta_T})], \\
    N_0(r) &= N_{\text{ref}} \exp[-\kappa_N \frac{a}{R_0} \Delta_N \tanh(\frac{(r-r_0)/a}{\Delta_N})],
\end{aligned}
\end{equation*}
with $\kappa_T = 6.9589$, $\kappa_N = 2.232$, $\Delta_T = \Delta_N = 0.3$, $r_0 = 0.5a$, and electron-to-ion temperature ratio $\tau = T_\text{e}/T_\text{i} = 1$.
The flat-profile case typically uses ITER-like parameters~\cite{mantica2020progress} with reduced temperature and density gradients.
Safety factor profiles $q(r)$ are prescribed to provide the desired magnetic shear $s(r) = (r/q)\mathrm{d}q/\mathrm{d}r$ in the core region.
Extended Data Fig.~\ref{fig:profile_linear} shows typical profiles.

\subsection{Linear eigenmode analysis.}
For each equilibrium, the most unstable linear ITG eigenmodes are obtained by running the NLT code in the linear regime and identifying the dominant toroidal mode number $n$.
The poloidal harmonic amplitudes $\phi_m(r)$ are extracted.
Real frequencies $\omega_r$ and growth rates $\gamma_L$ are computed from the time evolution of the electrostatic potential at different radial positions, confirming the coherence of each eigenmode.
The full width at half maximum of a poloidal harmonic $\Delta_{\text{mph}}$ is obtained by fitting the harmonic amplitude envelope $|\phi_m(r)|$ to a Gaussian; the measured width is nearly independent of the specific fitting procedure.

\begin{itemize}
\item Data availability: The data underlying all figures and Extended Data figures of this study, together with the analysis and plotting scripts, are openly available in <repository, e.g. Zenodo> at <DOI or URL>. The raw three-dimensional simulation output files are too large to be deposited and are available from the corresponding author upon reasonable request.
\item Code availability: The NLT code used in this study is available in <repository, e.g. GitHub/Zenodo> at <DOI or URL>, version <x.y>.
\end{itemize}

\section*{Extended Data}

\renewcommand{\figurename}{Extended Data Fig.}
\setcounter{figure}{0}

\begin{figure}[H]
    \centering
    \includegraphics[scale=1]{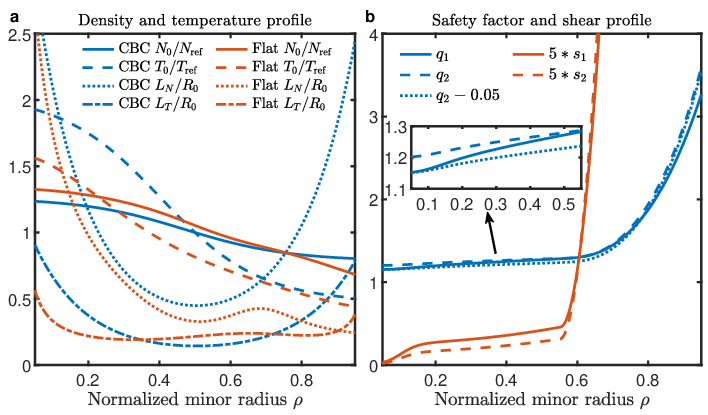}
    \caption{\textbf{Equilibrium profiles in linear simulations.} \textbf{a}, normalized density (solid), temperature (dashed), density scale length (dotted), and temperature scale length (dash-dotted). Blue: Cyclone base case; orange: ITER-like flat case. \textbf{b}, Safety factor (blue) and magnetic shear, $5s$, (orange); $s_1 \approx 0.08$ for $q_1$; $s_2 \approx 0.05$ for $q_2$.}
    \label{fig:profile_linear}
\end{figure}

\begin{figure}[H]
    \centering
    \includegraphics[scale=1]{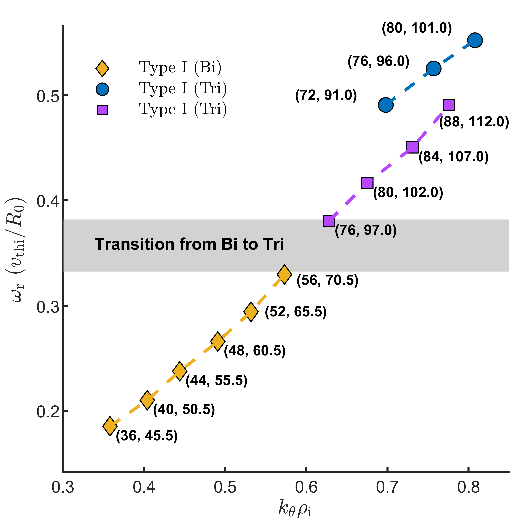}
    \caption{\textbf{Numerical calibration of $\alpha_{\text{PLD}}$.} Real frequency $\omega_r$ of Type I modes versus poloidal wavenumber $k_\theta\rho_\text{i}$ for a weak-shear JET case with Cyclone-like profiles and the $q_1$ profile (Extended Data Fig.~\ref{fig:profile_linear}). Index pairs are labelled as $(n,m)$; the two Tri-harmonic series are plotted as blue circles and purple squares. At small $k_\theta\rho_\text{i}$ (hence small $\omega_r$), the modes are Type I (Bi); as $k_\theta\rho_\text{i}$ increases, $\omega_r$ rises and the modes transition to Type I (Tri). The Bi--Tri transition occurs at $\omega_{r,\text{lim}}\approx 0.39\,v_{\text{th,i}}/R_0$. With $q \approx 1.275$ and $\omega_\text{t} \approx v_{\text{th,i}}/(qR_0)$, the Bi--Tri transition condition gives $\alpha_{\text{PLD}} = \omega_\text{t}/\omega_{r,\text{lim}} \approx 2.0$.}
    \label{fig:alpha_PLD}
\end{figure}

\begin{figure}[H]
    \centering
    \includegraphics[scale=1]{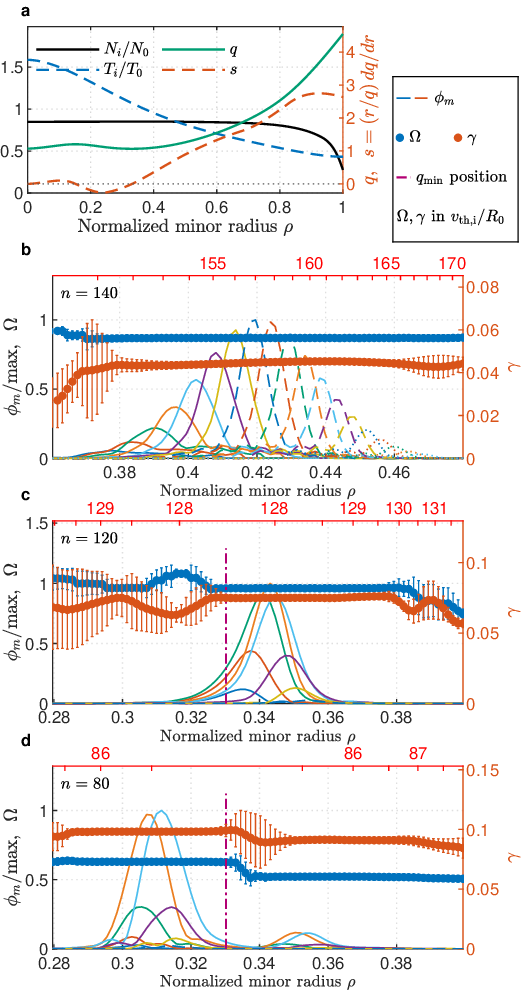}
    \caption{\textbf{Test of the critical shear for a weakly reversed magnetic shear configuration.} \textbf{a}, Equilibrium profiles for the ITER-parameter case: normalized ion density $N_i/N_0$, normalized ion temperature $T_i/T_0$, safety factor $q$ and magnetic shear $s=(r/q)\,\mathrm{d}q/\mathrm{d}r$. The core shear is weakly reversed, $s<0$ over the inner region, and $q$ reaches its minimum near $\rho\approx0.3$. \textbf{b-d}, Poloidal harmonic amplitudes $\phi_m$ (coloured curves, left axis), together with the real frequency $\Omega$ (blue circles) and the growth rate $\gamma$ (orange circles), in units of $v_{\text{th,i}}/R_0$, as functions of the normalized minor radius $\rho$, for $n=140$, $120$ and $80$. The three toroidal mode numbers are unstable in different radial regions and organize into the three structures distinguished by the isotropy criterion. \textbf{b}, $n=140$: many harmonics spread over many rational surfaces, forming a radially extended Type II mode. \textbf{c}, $n=120$: three harmonics of comparable amplitude, $m=127$, $128$ and $129$, share a common localization centre, forming a Type I (Tri-harmonic) mode. \textbf{d}, $n=80$: two dominant harmonics, $m=85$ and $86$, share a common localization centre, forming a Type I (Bi-harmonic) mode; its real frequency is lower than that of the Tri-harmonic mode, as expected at the Bi--Tri transition.}
    \label{fig:linear_reversed}
\end{figure}

\begin{figure}[H]
    \centering
    \includegraphics[scale=1]{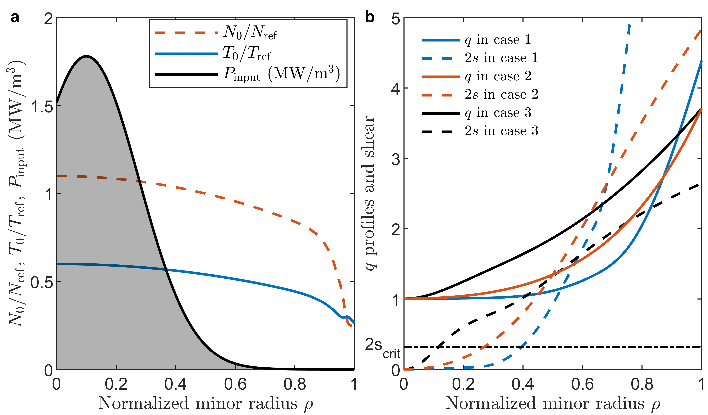}
    \caption{\textbf{Equilibrium profiles in nonlinear simulations.} \textbf{a}, Normalized density (orange dashed), temperature (blue solid) and heating power deposition (black shading) profiles for DIII-D parameters. The total heating power is $2.5~\text{MW}$, with $N_{\text{ref}}=2.0\times10^{19}~{\rm m}^{-3}$ and $T_{\text{ref}}=3.0~{\rm keV}$. \textbf{b}, Safety factor (solid) and magnetic shear, $2s$, (dashed) for the case 1 (finite-shear, black), case 2 (orange) and case 3 (weak-shear, blue) cases; $q(\rho=0)=1.004$. The horizontal black dotted line marks the weak-magnetic-shear threshold $s_{\text{crit}}$ predicted by the theory.}
    \label{fig:profile_nonlinear}
\end{figure}

\begin{figure}[H]
    \centering
    \includegraphics[width=\textwidth]{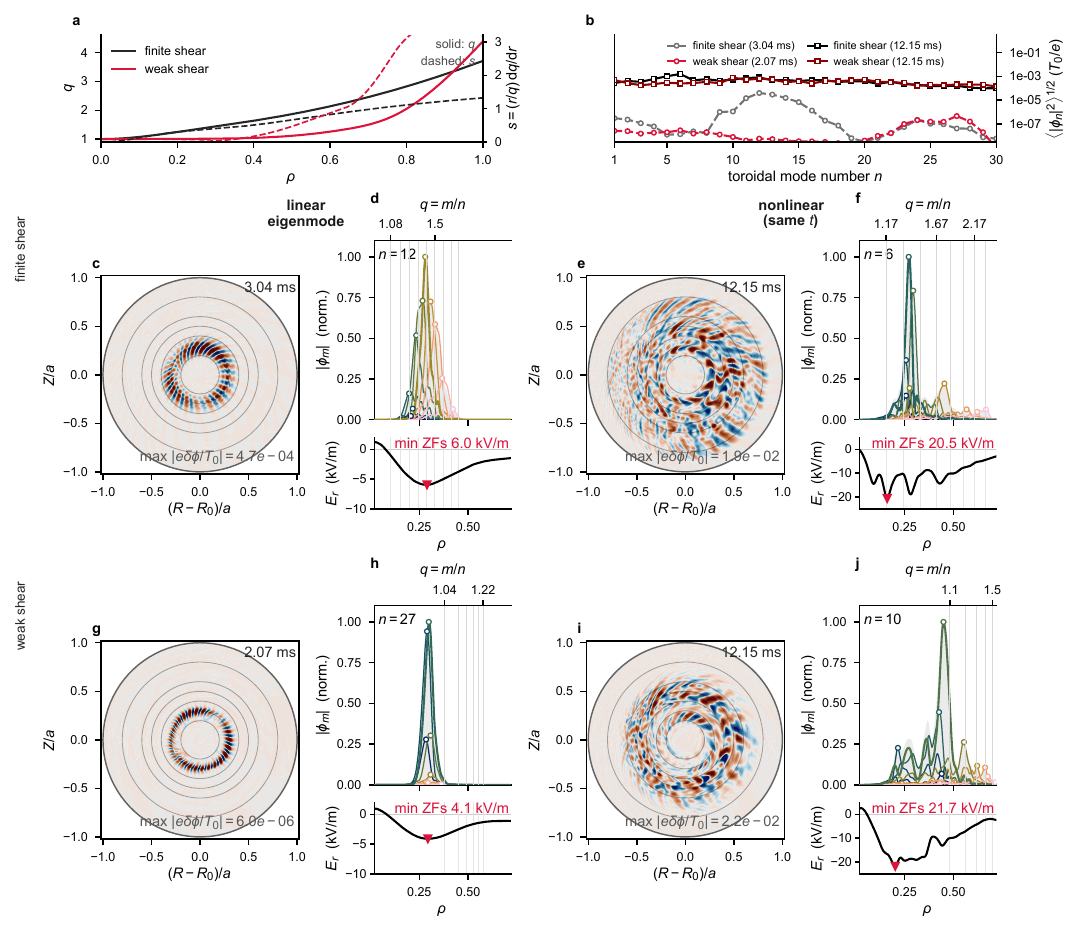}
    \caption{\textbf{Core magnetic shear controls the poloidal and toroidal structure of ion-scale turbulence.} Global gyrokinetic simulations of two DIII-D-like circular equilibria ($R_0=1.67$~m, $a=0.60$~m, $B_0=1.9$~T, $T_0=3$~keV, $n_0=2\times10^{19}$~m$^{-3}$, adiabatic electrons) that differ only in the core safety factor. \textbf{a}, Safety factor $q$ (solid, left axis) and magnetic shear $s=(\rho/q)\,\mathrm{d}q/\mathrm{d}\rho$ (dashed, right axis) versus normalized minor radius $\rho$. \textbf{b}, Volume r.m.s. amplitude of each toroidal harmonic (circles, quasi-linear; squares, nonlinear; black, finite; red, weak). \textbf{c}--\textbf{f}, finite magnetic shear and \textbf{g}--\textbf{j}, weak magnetic shear, with quasi-linear (left) and nonlinear (right). \textbf{c,e,g,i}, the fluctuant potential $e\delta\phi/T_0$ on the poloidal cross-section; \textbf{d,f,h,j}, the poloidal harmonics $|\phi_m|$ of the largest-amplitude toroidal mode (rational surfaces $q=m/n$ on the top axis) together with the total zonal flows $E_r=-\mathrm{d}\langle\phi_z\rangle/\mathrm{d}\rho$ (red triangle, minimum) on a shared $\rho$ axis. Harmonics are normalized to their own maxima. The two nonlinear snapshots are taken at the same time, 12.15~ms. In the quasi-linear phase the finite-shear mode is a discrete toroidal mode ($n=12$) whose harmonics form a wide ballooning envelope peaking at successive rational surfaces, whereas the weak-shear mode ($n=27$) collapses onto two harmonics ($m=27/28$). In the nonlinear phase the radial envelopes broaden in both runs, and the minimum of $E_r$ is comparable in the two runs ($-20.5$ and $-21.7$~kV/m).}
    \label{fig:nonlinear_structure}
\end{figure}

\section*{Declarations}

\begin{itemize}
\item Funding: This work was supported by the National Natural Science Foundation of China under Grant Nos. 12535014 and 12505261, the Strategic Priority Research Program of the Chinese Academy of Sciences under Grant No. XDB0790201 and the National MCF Energy R\&D Program of China under Grant No. 2025YFE03000100.
\item Author contributions: Z. Wang carried out the numerical simulation and diagnosis. S. Wang conceptualized the micro-isotropy and the critical magnetic shear. Z. Wang, T. Wu and S. Wang analyzed the results and edited the manuscript.
\item Competing interests: The authors declare no competing interests.
\item Correspondence and requests for materials should be addressed to S. J. Wang.
\end{itemize}


\bibliography{ref}

\end{document}